\documentclass[conference, 9pt]{IEEEtran}
\IEEEoverridecommandlockouts
\usepackage{cite}
\usepackage{algorithmic}
\usepackage[linesnumbered,ruled,vlined]{algorithm2e}
\algsetup{linenosize=\small}
\usepackage{graphicx}
\usepackage{multirow}
\usepackage{textcomp}
\usepackage{xcolor}
\usepackage{scalerel}
\usepackage{graphics}
\usepackage[utf8]{inputenc}
\usepackage{amsfonts}
\usepackage[fleqn]{amsmath}
\usepackage[hidelinks]{hyperref}
\usepackage{array}
\usepackage{multirow}
\usepackage{booktabs}
\usepackage{amssymb}
\usepackage{color}
\usepackage{array}
\usepackage{hhline}
\usepackage{soul}
\usepackage{comment}
\usepackage{caption}
\usepackage{stfloats}
\def\BibTeX{{\rm B\kern-.05em{\sc i\kern-.025em b}\kern-.08em
    T\kern-.1667em\lower.7ex\hbox{E}\kern-.125emX}}
\begin{document}
\bstctlcite{IEEEexample:BSTcontrol}
\title{From FPGAs to Obfuscated eASICs: Design and Security Trade-offs\\[-1.4ex]
}


\author{Zain Ul Abideen, Tiago Diadami Perez, Samuel Pagliarini 
         \\
\IEEEauthorblockA{Centre for Hardware Security, Tallinn University of Technology (TalTech), Estonia\\ \{zain.abideen, tiago.perez, samuel.pagliarini\}@taltech.ee}\\[-7ex]
\thanks{This work was partially supported by the EC through the European Social Fund in the context of the project ``ICT programme". It was also partially supported by the Estonian Research Council grant MOBERC35.}
}



\maketitle

\begin{abstract}
Threats associated with the untrusted fabrication of integrated circuits (ICs) are numerous: piracy, overproduction, reverse engineering, hardware trojans, etc. 
The use of reconfigurable elements (i.e., look-up tables as in FPGAs) is a known obfuscation technique. In the extreme case, when the circuit is entirely implemented as an FPGA, no information is revealed to the adversary but at a high cost in area, power, and performance. In the opposite extreme, when the same circuit is implemented as an ASIC, best-in-class performance is obtained but security is compromised. This paper investigates an intermediate solution between these two. Our results are supported by a custom CAD tool that explores this FPGA-ASIC design space and enables a standard-cell based physical synthesis flow that is flexible and compatible with current design practices. Layouts are presented for obfuscated circuits in a 65nm commercial technology, demonstrating the attained obfuscation both graphically and quantitatively. Furthermore, our security analysis revealed that for truly hiding the circuit's intent (not only portions of its structure), the obfuscated design also has to chiefly resemble an FPGA: only some small amount of logic can be made static for an adversary to remain unaware of what the circuit does. 

\end{abstract}

\begin{IEEEkeywords}
Hardware Obfuscation, Secure ASIC Design, CAD, Reconfigurable obfuscation, Reverse engineering
\end{IEEEkeywords}

\section{Introduction} \label{sec:intro}

Shipment of semiconductor devices is forecast to surpass one trillion units in the year 2021, the third time this mark is surpassed in a calendar year since 2018 \cite{numberDevices}. The majority of those devices are being manufactured by foundries that subscribe to the fab-for-hire model. 
A rogue element within the foundry can mount fabrication-time attacks, i.e., the foundry and its employees are considered potential adversaries. Many potential threats regarding third-party foundries have been studied in recent years, include tampering, counterfeiting, reverse engineering, and overproduction \cite{ref_2_nist}. On the other hand, many techniques have been devised to mitigate threats from untrusted fabrication. 
Countermeasure techniques to increase the IC security against not only third-party foundries but also from the end-user have been recently demonstrated. Notable examples include IC Camouflaging \cite{cam_1, cam_2, cam_3}, Logic Locking \cite{logic_1, 7362173, logic_5}, and Split Manufacturing \cite{split_1, split_2}.

\begin{figure}[tb]
\centering \footnotesize
\includegraphics[width=1.0\linewidth]{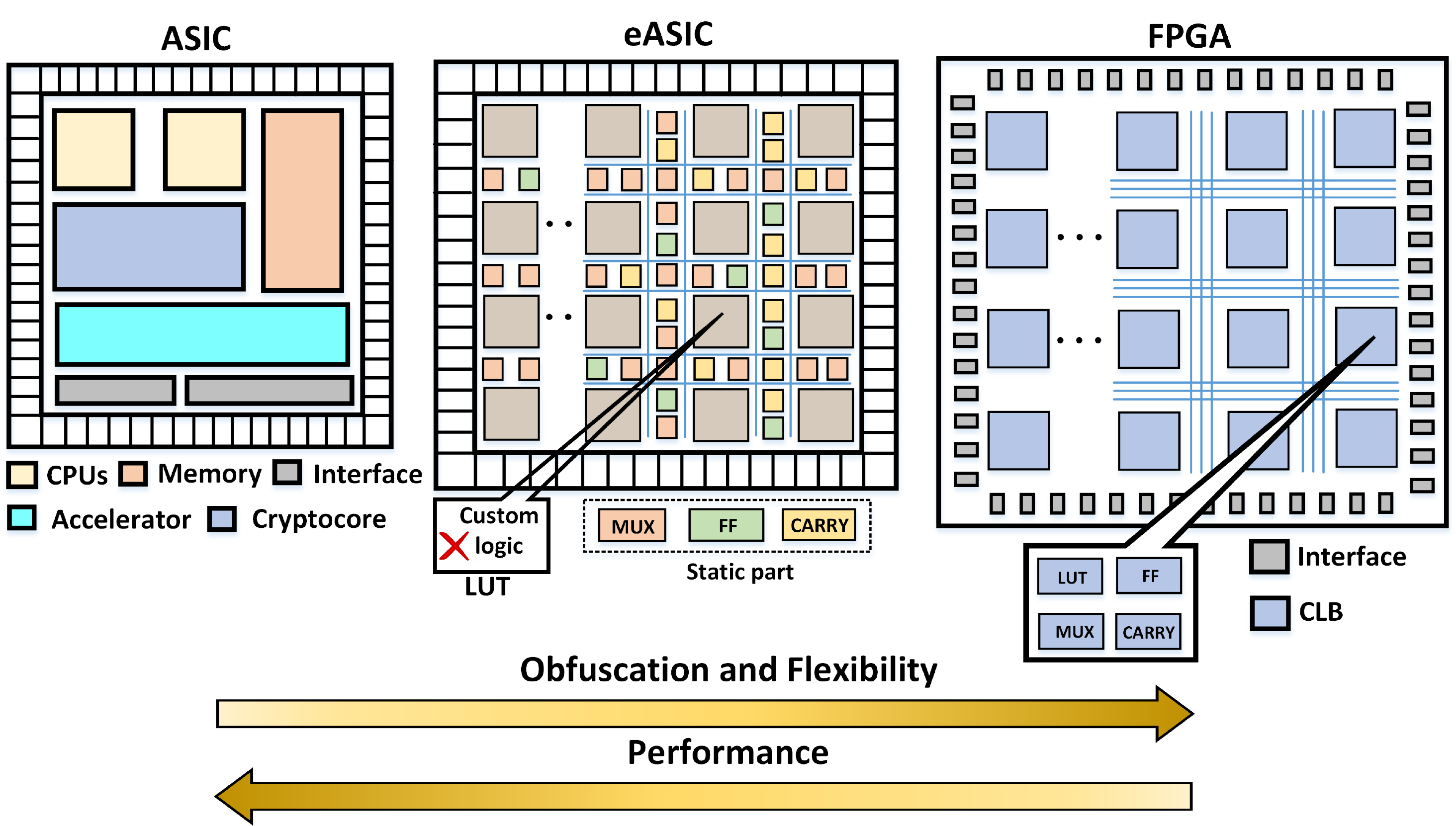}
\caption{The design obfuscation landscape, from ASICs to FPGAs. The relative sizes are notional.}
\vspace{-10pt}
\centering
\label{fig:motivation}
\end{figure}

 Generally speaking, all of the aforementioned countermeasures attempt to ``hide'' the design from adversaries and can be classified as obfuscation techniques. Unfortunately, none of these techniques is currently adopted in large-scale production of ICs, for reasons that include (lack of) practicality \cite{split_1} and insufficient security guarantees \cite{eval_logic}. Another approach towards obfuscation is the use of an FPGA (or FPGA-like) design, where the configuration \emph{bitstream serves as a key} to unlock the functionality of the circuit \cite{reconfigure_3}. Our paper too explores this possibility. 
The fabric in an FPGA contains reconfigurable elements, but this flexibility incurs a limited performance. On the other hand, ASIC requires one-time placement, it is static (non-reconfigurable), but it provides best-in-class performance. As shown in  Fig. \ref{fig:motivation}, performance increases if we move from right to left. Contrarily, area, obfuscation, and flexibility increase if we move from left to right. However, we argue that \emph{neither extremes of the spectrum are a good design point} for circuits that have stringent security and performance constraints. An intermediate solution is a better trade-off and this is precisely the motivation for our work. We term our in-between solution an ``embedded ASIC'' (\textbf{eASIC}). 


\textbf{Design obfuscation concept:} In this work, we propose to obfuscate a design by exploiting the best of both worlds. The generated device is a hybrid which includes \textit{reconfigurable elements} (analogous to the FPGA) and also includes ASIC cells as \textit{static elements}, i.e., gates with fixed functionality after fabrication. Previous research on obfuscation by reconfigurable elements has focused on keeping the reconfigurable portion as small as possible~\cite{lut2, lut3}, which is logical if the goal is to keep overheads under control. However, we later show that true hiding of the circuit's intent requires a \emph{high degree of obfuscation} that is usually not explored in the state of the art. Thus, our eASIC device is largely non-functional until it is programmed. Our main contribution is a tool for automatically obfuscating a design in the form of eASIC, where the obfuscation range can be from 0 to 100\%. Furthermore, its physical synthesis flow is standard-cell based that is compatible with current design and fabrication practices.

\section{A CAD flow for eASIC} \label{sec:cad_flow}

\begin{figure}[tb]
\centering 
\includegraphics[width=0.72\linewidth]{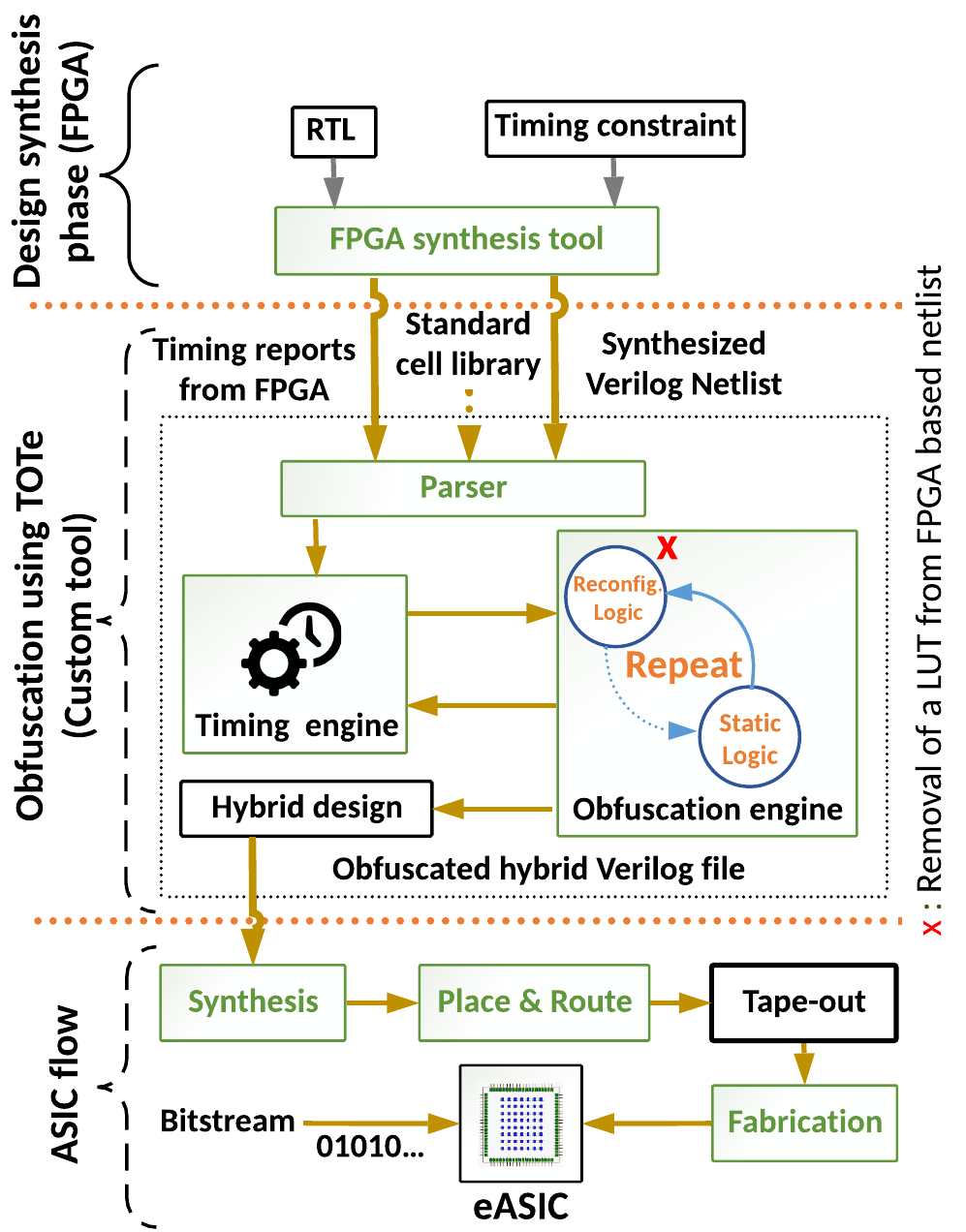}
\caption{The CAD flow for eASIC, combining FPGA and ASIC synthesis.}
\vspace{-10pt}
\centering
\label{fig:design_flow}
\end{figure}

Our CAD flow is centered around a tool named \textbf{T}uneable Design \textbf{O}bfuscation \textbf{T}echnique using \textbf{e}ASIC, or \textbf{TOTe} for short. This section explains the CAD flow of eASIC and TOTe's main features. 
TOTe generates a hybrid design with static and reconfigurable elements, which we refer to as eASIC. For the reconfigurable elements, we implement the logic utilizing the notion of programmable LUTs (Look Up Tables) - same as in FPGAs. The complete TOTe design flow for generating an eASIC is shown in Fig. \ref{fig:design_flow} and it consists of three phases.


In the \textbf{first phase} of our flow, the design under obfuscation, described in register-transfer level (RTL) form, is synthesized using a commercial FPGA synthesis tool. As a result, the netlist contains all typical FPGA primitives, i.e., FFs, MUXs, and LUTs. The input design requires no special annotations, synthesis pragmas, or any other change in its representation. 

Next, in the \textbf{second phase}, TOTe requires the ASIC standard cell library of choice. As highlighted in the center of Fig. \ref{fig:design_flow}, the core idea of TOTe is to \textbf{replace reconfigurable logic for static logic}. For TOTe, only the LUTs are treated as reconfigurable logic, and, any other primitives from the FPGA synthesis are automatically transformed into static logic. For this phase, the designer provides an obfuscation target in terms of percentage, which determines the portion of logic that will remain reconfigurable as LUTs. TOTe builds a tree representation of the circuit where primitive types are annotated for every instance. For LUTs, in particular, the tool also annotates their masking patterns (i.e., the portion of the bitstream associated with an individual LUT). By using truth tables populated by the masking patterns, TOTe builds combinational logic that is equivalent to the LUT's intended usage. This equivalent logic is what we refer to as the static logic that is used for replacing the reconfigurable logic. Notice that the equivalent logic is not a hardwired configurable LUT. 



TOTe utilizes its own \textbf{obfuscation and timing engines}. These engines drive the security vs. performance trade-offs of the tool. Algorithm \ref{alg:obfuscation_engine} highlights the main loop of the tool and its different operations. Here, $L$ is a list of LUTs, $P$ is a list of timed paths, and $obf_c$ is the obfuscation target criterion. The variables $L\textsubscript{ST}$ and $L\textsubscript{RE}$ are lists of LUTs in static and reconfigurable form, respectively. The obfuscation engine is executed until the desired number (determined by the obfuscation target) of LUTs has been transformed into static logic (line 3), where the SIZE\_OF function returns the size of a list. Inside the obfuscation inner loop, the critical path is identified (line 4) using the FIND\_CRITICAL function, then the slowest LUT on that path is identified using the FIND\_SLOWEST function (line 5). If the identified LUT is a reconfigurable LUT (line 6), the lists of LUTs are updated (lines 7-8) and the timing engine recalculates the timing of affected paths (line 9). If the identified LUT is not a reconfigurable LUT (line 10), the path is removed (line 11) and the loop continues (line 3). The INSERT and REMOVE functions update the lists as hinted by their names. The function GEN\_CASE\_0\_1 generates the constraints to force the values of the LUTs inputs for the timing and power analysis during physical synthesis, otherwise, each LUT would be timed for its worst timing arc instead of the real implemented timing arc when the LUT is programmed. The algorithm merges $L\textsubscript{ST}$ and $L\textsubscript{RE}$ to generate eASIC and returns. Finally, TOTe exports an obfuscated hybrid Verilog file (eASIC), timing report, and area report. A designer can repeat this procedure until he achieves his obfuscation and performance targets.

\begin{algorithm}[h]
\small
\caption{\textbf{T}uneable \textbf{O}bfuscation \textbf{T}echnique using \textbf{e}ASIC}
\label{alg:obfuscation_engine}
  \SetKwInOut{Require}{Require}
  \SetKwInOut{Ensure}{Ensure}
  \SetKwInput{KwInput}{Input}
  \SetKwInput{KwOutput}{Output}
\DontPrintSemicolon
  \KwInput{$L$ $(list$ $of$ $LUTs$), $P$ $(list$ $of$ $paths)$,  $obf_c$ $(obfuscation$ $criterion)$}
  \KwOutput{$eASIC \gets f(input)$}
  
  $L\textsubscript{ST} \gets \phi$
  
  $L\textsubscript{RE} \gets L$
  
    
  \While{SIZE\_OF {($L\textsubscript{ST}$)} $\leq$ $obf_c$}
  {
  $path$ $\gets$ FIND\_CRITICAL($P$)
  
  $lut$ $\gets$ FIND\_SLOWEST($path$)
  
      \If{$lut \in L\textsubscript{RE}$}
        {
        INSERT($lut$, $L\textsubscript{ST}$)
        
        REMOVE($lut$, $L\textsubscript{RE}$)
        
        UPDATE\_TIMING($lut$, $P$)
        }
        \Else{
        REMOVE($path$, $P$)
        }
}
    
      \For{each $lut \in L\textsubscript{RE}$}
    {
    GEN\_CASE\_0\_1($lut$)
    }

$eASIC \gets L\textsubscript{ST} \cup L\textsubscript{RE}$
\end{algorithm}

In the \textbf{third and final phase}, the obfuscated netlist from TOTe is synthesized using any commercial ASIC CAD tool and implemented using an also commercial physical synthesis tool where traditional P\&R, CTS, DRC, LVS, etc. steps are executed. Finally, the tapeout database is sent to the foundry for fabrication. Once the fabricated parts are delivered, they have to be programmed (i.e., using a bitstream as in FGPAs) for the eASIC design to be completed and functional. 

\section{Experimental results using TOTe} \label{sec:results}

This section reports the analysis of security versus performance, security versus area for selected designs and reports the results for numerous designs after obfuscation. For all experimental results that follow, FPGA synthesis was executed in Vivado and the targeted device is Kintex-7 XC7K325T-2FFG900C, which contains only 6-input LUTs. 
For the ASIC flow, the implementations are done using a commercial 65nm PDK with three standard cell flavors (LVT/SVT/HVT) and tools from Cadence suite (i.e., Genus and Innovus). However, we emphasize that TOTe is \textbf{agnostic with respect to PDKs, libraries, and tools}. In the experiments that follow, we prevent the FGPA synthesis from inferring any BRAM or instantiating any DSP cells. This choice makes for very clear trade-offs when obfuscating the logic. Nevertheless, if a given design requires memory cells, TOTe has the capability to translate the inferred BRAMs into compiled SRAMs for the specific ASIC technology utilized.


\subsection{Custom standard-cell based LUTs} \label{subsec:lut_implementation}
\begin{figure}[tb]
\centering \footnotesize
\includegraphics[width=0.8\linewidth]{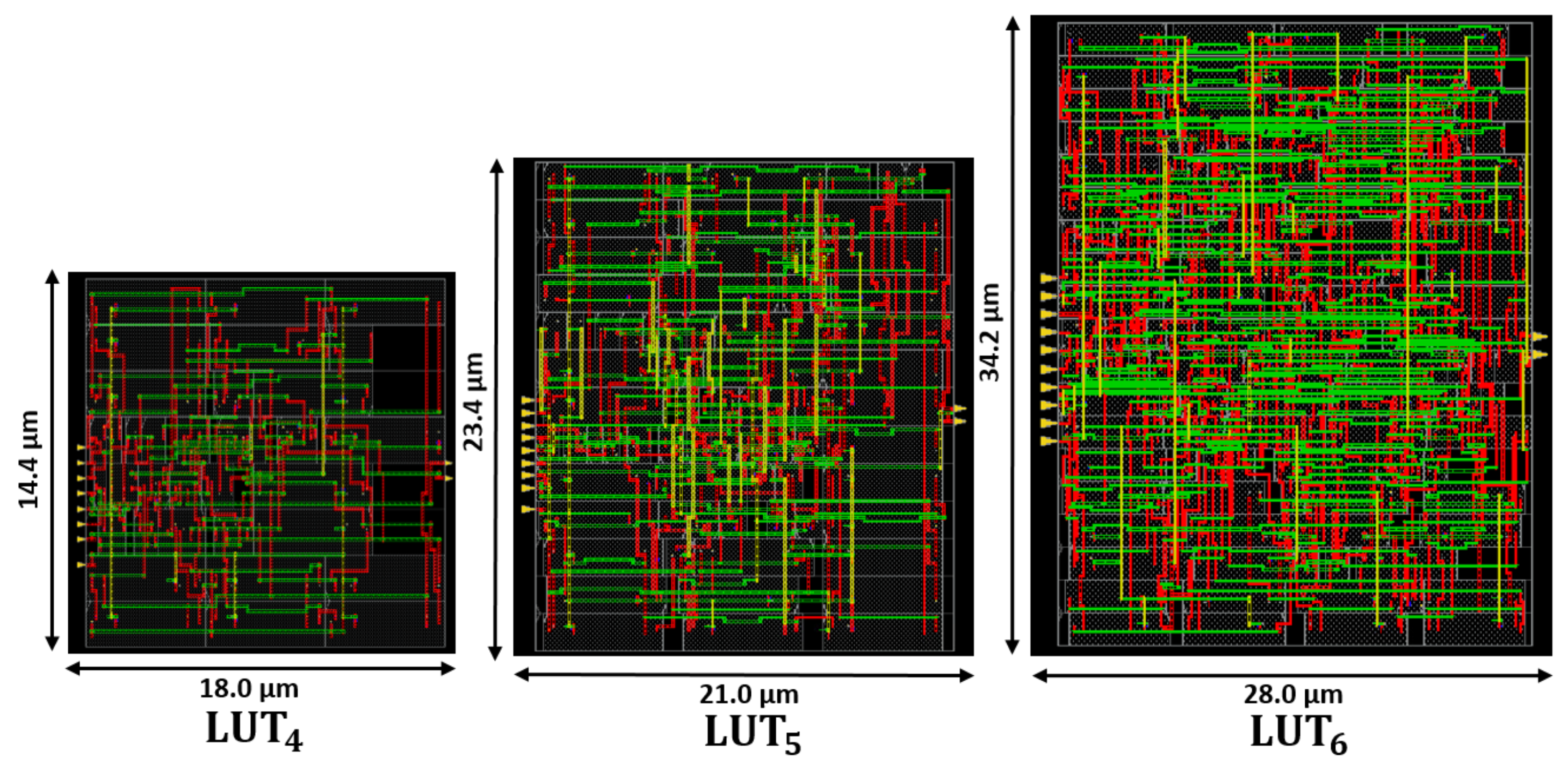}
\caption{The layout of macros for LUT\textsubscript{4}, LUT\textsubscript{5} and LUT\textsubscript{6}. Implementation was executed in Cadence Innovus.}
\vspace{-10pt}
\centering
\label{fig:lut_imp}
\end{figure}

The premise of eASIC is to have reconfigurable and static elements that can be integrated transparently. For this reason, we have designed our own custom LUTs (LUT\textsubscript{1}, LUT\textsubscript{2},..., LUT\textsubscript{6}) by following VPR's template \cite{vpr8}. Different from FPGAs that generally implement only one LUT size, for eASIC we have the flexibility to implement more than one size because our design intent will not change. By doing this, we preserve area and potentially increase the performance of eASIC. The layouts for LUT\textsubscript{4}, LUT\textsubscript{5}, and LUT\textsubscript{6} macros are shown in Fig. \ref{fig:lut_imp}. These blocks are highly compact since the main design goal for them was area/density. 
Each LUT has its own registers for storing the configuration, a functionality that is enabled by including three extra configuration pins ($serial\_in$, $serial\_out$, and $enable$). The LUTs are connected to one another in a daisy chain that is analogous to a scan chain. We chose a flip-flop based implementation to make our framework easily portable between technologies, and also, make the floorplan and placement almost effortless. Using SRAM for storing those bits, on the other hand, can save area and power but requires extra effort during implementation since memories need special power routing and have to be strategically placed to achieve the best performance. 

\subsection{Design Space Exploration in TOTe} \label{subsec:design_space}

For our first experiment, we selected a small but representative design which covers all possible FPGA primitives: a schoolbook multiplier (SBM), which is a bit-serial polynomial multiplication circuit. 
For a SBM design that is synthesized targeting a very high frequency, the $CP$ and $sumCP$ become, as calculated by TOTe's timing engine, 0.490 ns and 16088.69 ns, respectively. These values correspond to a design obfuscated at 100\%, i.e., the design has only reconfigurable logic. The absolute accuracy of these values is not relevant since final timing analysis is performed using a commercial physical synthesis engine later. 



\begin{figure}[tb]
\centering
\includegraphics[width=0.95\linewidth]{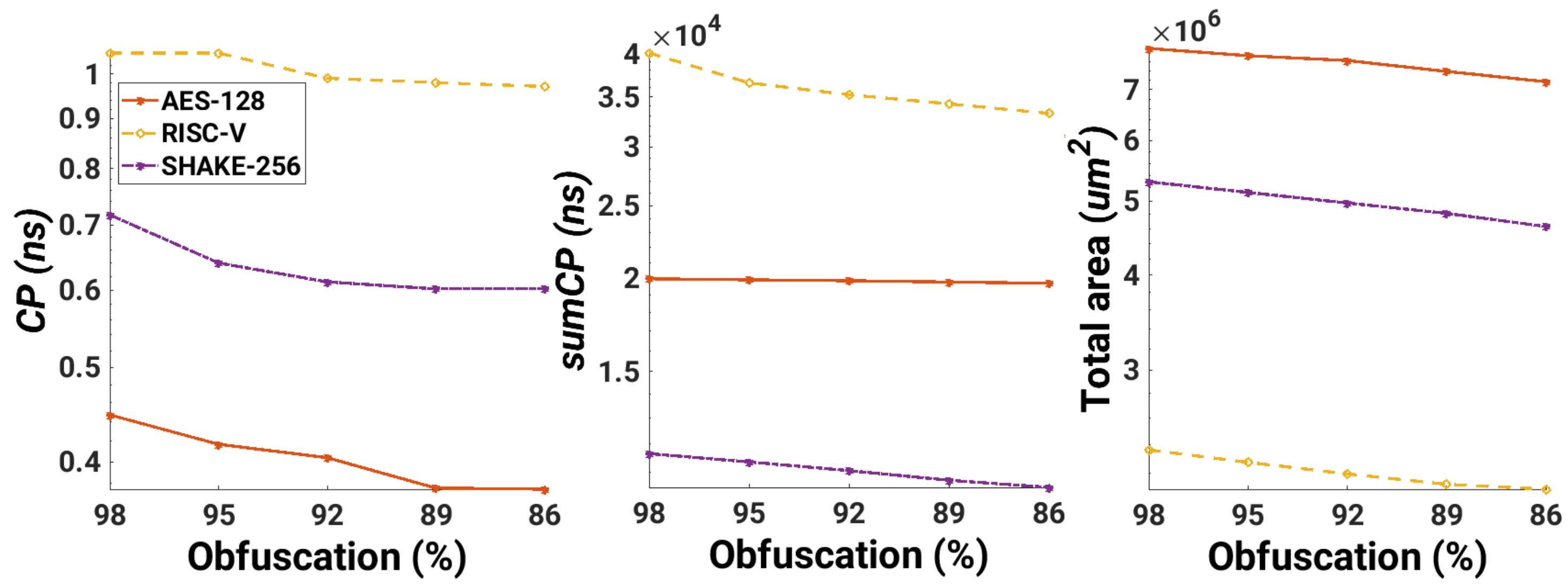}
\caption{Obfuscation results for AES-128, RISC-V and SHAKE-256 using TOTe.}
\vspace{-10pt}
\centering
\label{fig:obfuscation_results}
\end{figure}

While the SBM design is an interesting motivational example, it showed that $CP$ tends to saturate while the $sumCP$ continues to improve as the obfuscation is reduced. Next, we wanted to determine if the same saturation trend appears for other designs and the results are reported in Table~\ref{tab:obfuscation_designs}. From these experiments, it is possible to conclude that the performance of numerous designs saturates incredibly fast as we decrease the obfuscation level, even when the obfuscation range is limited to 86-100\%. Moreover, the results for AES-128~\cite{aes_128}, RISC-V, and SHAKE-256 have been depicted in Fig.~\ref{fig:obfuscation_results}. Several other designs, including ISCAS'85 benchmarks and known opencores, have been evaluated. The complete set of results can be found in our git repository \cite{git_results}. Concerning runtime, TOTe requires only a few minutes for big designs. Then, commercial CAD tools require a considerable amount of time for logic and physical synthesis (same as conventional ASIC flow). In summary, TOTe does not become a bottleneck in the design flow.  


\begingroup
\setlength{\tabcolsep}{2.0pt} 
\begin{table} [t]
\footnotesize \centering
\caption{Detailed results for selected designs using TOTe}
\label{tab:obfuscation_designs}
\begin{tabular}{|p{1.18cm}|p{0.515cm}|p{1.2cm}|p{0.7cm}|p{1.35cm}|p{1.18cm}|p{0.75cm}|p{0.75cm}|}
\hline 
\textbf{Design} & \textbf{Obf. (\%)} & \textbf{\textit{sumCP} \, ($ns$)} & \textbf{\textit{CP} ($ns$)} & \textbf{Area-RE ($\mu m^2$)} & \textbf{Area-ST ($\mu m^2$)} & \textbf{LUT\, (RE)} & \textbf{LUT \, (ST)} \\ \hline \hline
\multirow{5}{*}{SBM} & 98 & 16088.690 & 0.490 & 13190.04 & 0 & 29 & 0  \\ \hhline{~-------}
& 95 & 15895.826 & 0.484 & 12762.00 & 21.40 & 28 & 1 \\ \hhline{~-------}
& 92 & 15877.962 & 0.464 & 12547.80 & 32.11 & 27 & 2 \\ \hhline{~-------}
& 89 & 15458.506 & 0.461 & 12438.72 & 37.56 & 26 & 3 \\ \hhline{~-------}
& 86 & 15370.682 & 0.459 & 12224.52 & 48.27 & 25 & 4 \\ \hline \hline
\multirow{5}{*}{PID} & 98 & 2547.581 & 0.756 & 445590.0 & 2816.82 & 896 & 18  \\ \hhline{~-------}
& 95 & 2466.254 & 0.642 & 432340.92 & 9441.36 & 869 & 45 \\ \hhline{~-------}
& 92 & 2391.963 & 0.592 & 421365.95 & 14928.84 & 841 & 73 \\ \hhline{~-------}
& 89 & 2348.613 & 0.568 & 407273.76 & 21974.94 & 814 & 100 \\ \hhline{~-------}
& 86 & 2322.462 & 0.543 & 392345.64 & 29439.00 & 787 & 127 \\ \hline \hline
\multirow{5}{*}{\parbox{1.18cm}{SHA-256 }} & 98 & 7425.731 & 0.962 & 1313150.76 & 10291.86 & 2195 & 44  \\ \hhline{~-------}
& 95 & 7354.593 & 0.871 & 1275984.00 & 28875.24 & 2128 & 111 \\ \hhline{~-------}
& 92 & 7322.155 & 0.871 & 1233448.56 & 50142.96 & 2060 & 179 \\ \hhline{~-------}
& 89 & 7301.945 & 0.871 & 1179674.64 & 77029.92 & 1992 & 246 \\ \hhline{~-------}
& 86 & 7164.025 & 0.871 & 1125799.56 & 103967.46 & 1925 & 313 \\ \hline \hline
\multirow{5}{*}{FPU} & 98 & 2909.063 & 0.707 & 1031676.84 & 1250.028 & 2487 & 50  \\ \hhline{~-------}
& 95 & 2734.008 & 0.650 & 1003225.68 & 2672.586 & 2412 & 126 \\ \hhline{~-------}
& 92 & 2572.952 & 0.650 & 966715.20 & 4498.11 & 2336 & 202 \\ \hhline{~-------}
& 89 & 2478.732 & 0.650 & 935060.04 & 6080.868 & 2259 & 279 \\ \hhline{~-------}
& 86 & 2410.211 & 0.650 & 893005.56 & 8183.592 & 2183 & 355 \\ \hline
\end{tabular}
\end{table}
\endgroup


\section{Security Analysis} \label{sec:security_analysis}

As compared to conventional logic locking, the LUTs introduced in eASIC are the key-gate equivalents. In principle, a single LUT\textsubscript{n} ought to be equivalent to $2^n$ XOR/XNOR key-gates. In practice, the LUT logic has similarities to a run of key-gates (see \cite{7362173}) due to the $n$-to-1 multiplexing nature of it, which reduces the search space and may possibly make eASIC vulnerable to well-known oracle-based attacks (e.g., SAT). However, notice that we are considering designs with target obfuscation rates higher than 86\%, which results in bitstreams with thousands of bits. Even for a small and combinational design as the ISCAS'85 c7552, 50\% of obfuscation requires a bitstream with approximately 11k bits. The SAT attack is not able to find the correct key, even running for more than 60 hours. On the other hand, the same circuit with 10\% of obfuscation and a bitstream in the order of 1k bits, the SAT attack was successful, requiring less than a minute to retrieve the correct key. In general, the high obfuscation percentages obtained in eASIC discourage an adversary from performing SAT attacks. Similar findings were reported in \cite{lut1,lut2,lut3}. However, our proposed eASIC design \emph{potentially} creates attack vectors that other approaches do not since a portion of the design is exposed (from an adversary point of view). We will focus on these attacks in the text that follows.



\subsection{Threat Model} \label{subsec:threat_model}

In our considered threat model, the primary adversary is the \emph{untrusted foundry}. We make no distinction whether the adversary is institutional or a rogue employee. Assuming the security of an eASIC design is a function of its static logic (fully exposed) and reconfigurable logic (protected by a bitstream that serves as a key), we make the following assumptions:

\begin{itemize}
    \item The main adversarial goal is to reverse engineer the design in order to pirate its IPs, overproduce the IC, or even to insert sophisticated hardware trojans. For this goal, the adversary \textbf{must} recreate the bistream.
    
    \item The adversary goal might also be to identify the circuit intent, even in the presence of obfuscation. For this goal, the adversary \textbf{does not need} to recreate the bistream.
        
    \item The adversary has access to the GDSII file of the eASIC design. He or she is skilled in IC design and has no difficulty in understanding this layout representation. The adversary enjoys access to state-of-the-art CAD tools for this end.
    
    \item The attacker can recognize the standard cells, thus the gate-level netlist of the obfuscated circuit can be easily recovered~\cite{reverse_netlist}.
    
    \item We assume that the attacker can differentiate between design inputs and reconfiguration pins \cite{eval_logic, yasin_logic_locking}, thus being able to effortlessly identify all LUTs and their programming order.
    
    \item We assume the adversary can group the standard cells present in the static logic and convert them back into reconfigurable logic (i.e., LUT representation)\footnote{This is a very generous concession since the static logic is repeatedly optimized during logic and physical synthesis. Nevertheless, we err on the side of caution and assume the adversary can achieve a perfect reconstruction of LUTs, which by itself is a reverse engineering problem.}.
    


    
    
    
\end{itemize}

In order to evaluate the security hardness of eASIC, we propose two different attacks: one based on the \emph{structure} of design and another based on the \emph{composition} of known different circuits. 
Being so, we believe that the adversary can learn and extract information by exploiting the static portion of the design, including: (1) the frequency of masking patterns (2) the composition of different designs.



\subsection{Structural Analysis Attack} \label{subsec:structural_analysis}
\textbf{Goal}: by statistical analysis means, decrease the key search space and attempt to recover the bitstream.

We recall again that the obfuscation engine of TOTe utilizes six variants of LUTs (LUT\textsubscript{1}, LUT\textsubscript{2}, ..., LUT\textsubscript{6}) during the obfuscation phase. However, the majority of the LUTs are LUT\textsubscript{6} due to the packing algorithm executed during FPGA implementation. Therefore, let us assume without loss of generality, that any FPGA-synthesized circuit contains only instances of LUT\textsubscript{6}s for our security analysis. 

\begin{figure}[tb]
\centering 
\includegraphics[width=0.55\linewidth]{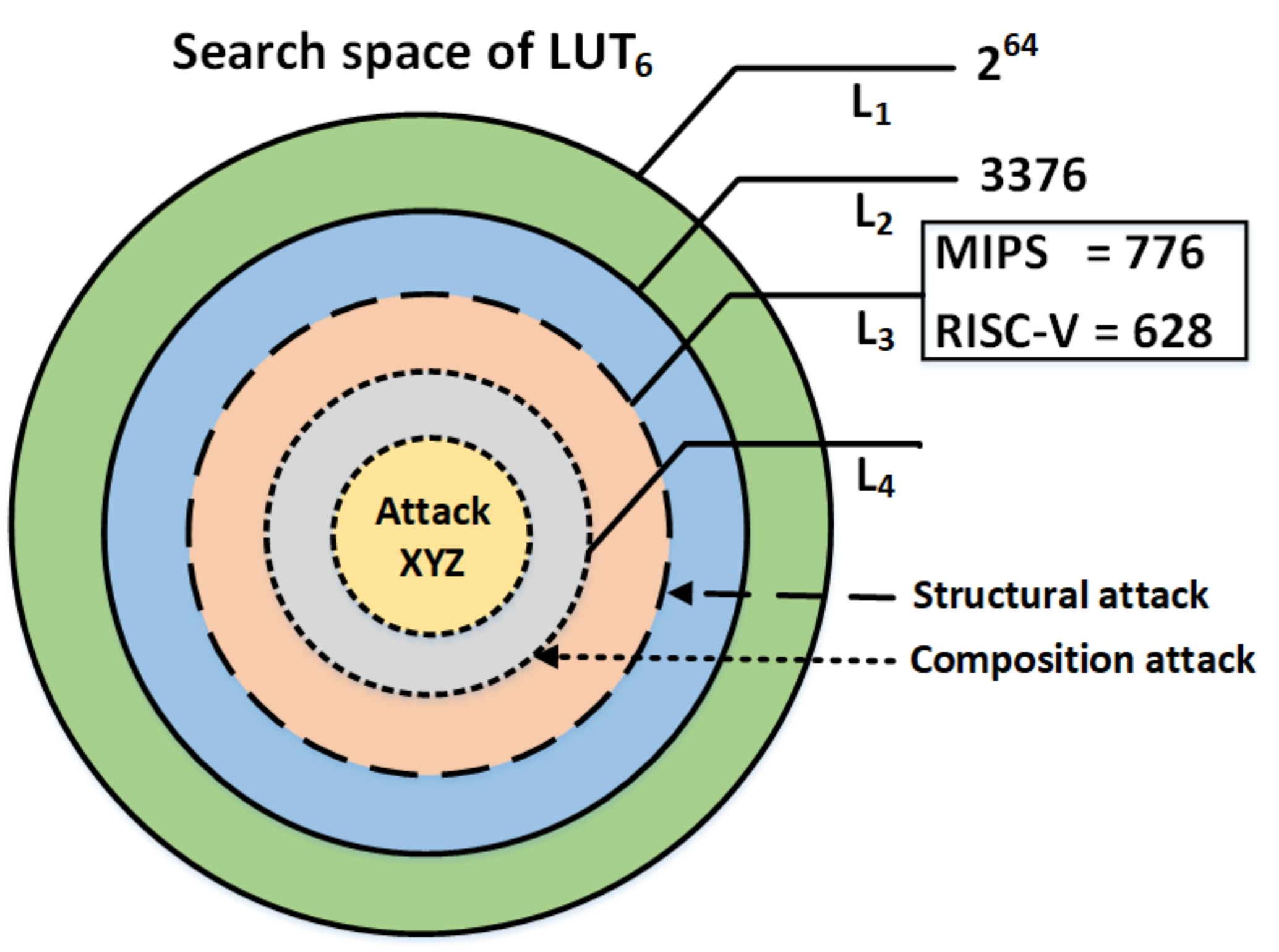}
\caption{The search space of LUT\textsubscript{6} as it shrinks with different attacks.}
\vspace{-10pt}
\label{fig:search_space}
\end{figure}

As we mentioned before, the key search space is $2^{64}$ for a single LUT\textsubscript{6}. But this assumption only holds if the FPGA synthesis is actually capable of exercising the entire key search space, which our results reveal that is far from possible. We have synthesized a large number of representative designs ($>$30) and counted how many unique LUT\textsubscript{6} masking patterns appear in the corresponding netlists. Designs of varied complexity, size, and functionality where added until the combined number of unique masking patterns appears to settle, forming a set of $m = 3376$ elements. This result alone, albeit being empirical, reduces the global search space from  L\textsubscript{1} to L\textsubscript{2} as illustrated in Fig.~\ref{fig:search_space}.

\begin{figure}[t]
\centering \footnotesize
\includegraphics[width=0.8\linewidth]{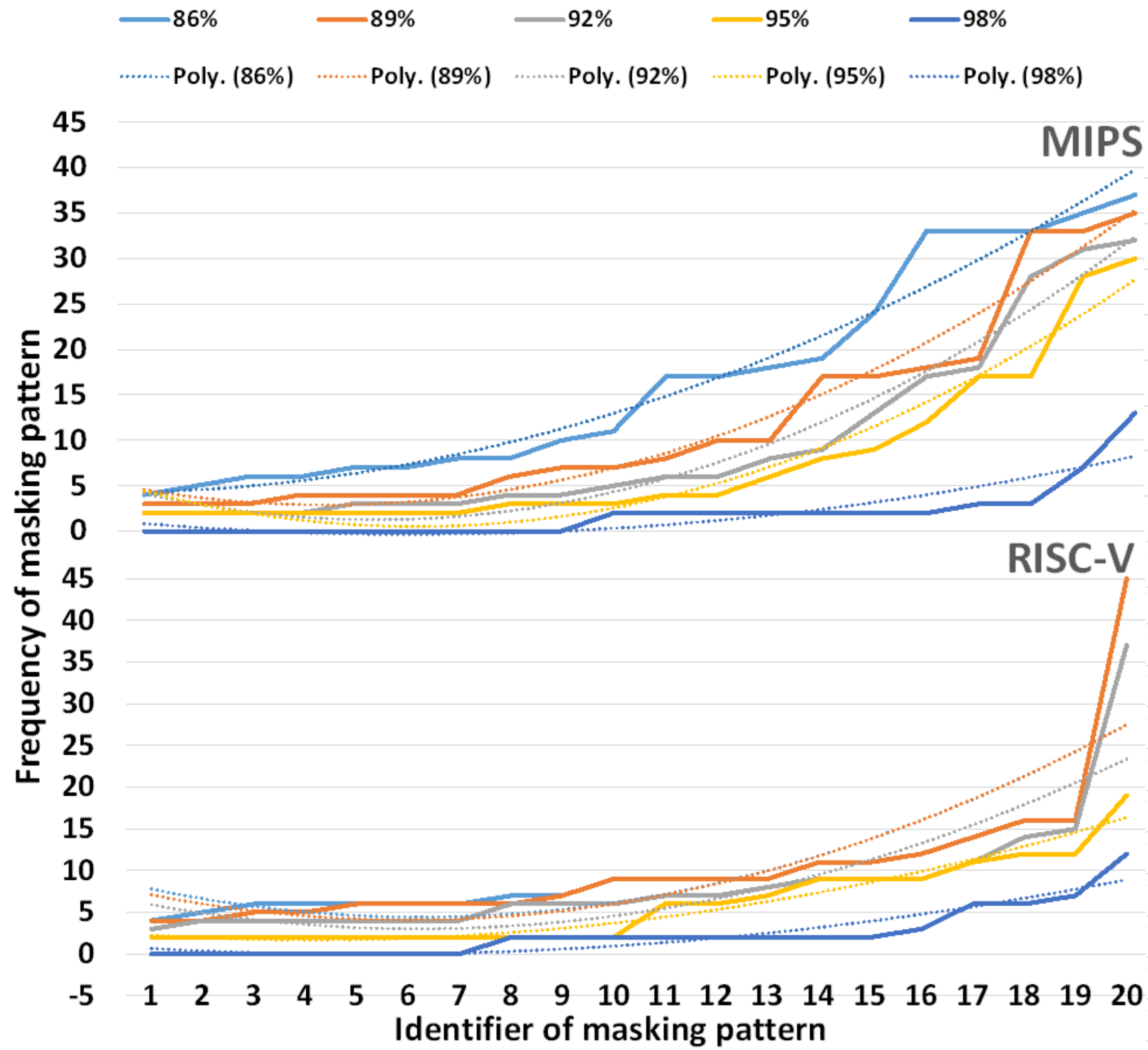}
\caption{The structural analysis of MIPS and RISC-V.}
\vspace{-10pt}
\centering
\label{fig:obfuscation_distr}
\end{figure}

In the next step, we targeted two processor designs in our statistical analysis: MIPS and RISC-V. For each circuit, we utilize tuples of $\langle pattern, frequency \rangle$ for tracking how often masking patterns repeat. The tuples are referenced by integer identifiers and ordered by frequency. Our analysis reveals that the MIPS netlist has 776 unique LUTs and that there are very few outliers that occur more than 50 times. Similarly, for RISC-V, there are 628 unique LUTs and only 3 occur more than 100 times. 
In practice, if the attacker could know for a fact that the obfuscated circuits are indeed MIPS and RISC-V, the search space would shrink further. The shrunk search spaces are labeled L\textsubscript{3} in Fig.~\ref{fig:search_space}. 
The question then becomes whether the static portion of the circuit is large enough for the adversary to be confident that the circuit under attack can be labelled as circuit C\textsubscript{1}, C\textsubscript{2}, or C\textsubscript{n}. We investigate this possibility by further analysing the behaviour of the frequency of masking patterns, as depicted in Fig.~\ref{fig:obfuscation_distr}. For this, we utilized polynomial trendlines for a portion of identifier of masking pattern, considering netlists generated by TOTe at 98\%, 95\%, 92\%, 89\%, and 86\% obfuscation levels. It is noteworthy that the trendlines become better frequency predictors as the obfuscation level is decreased. For RISC-V, in particular, the adversary can guess a small number outliers and the best guess (when obfuscation is 86\%) is far from the original frequencies ($>$100).




%
%
%


\subsection{Composition Analysis Attack} \label{subsec:composition_analysis}

\textbf{Goal}: identify the circuit by correlation to known circuits.

This attack also exploits the frequency of the LUT\textsubscript{6}, but here we correlate entire designs (instead of pattern-frequency tuples) based on their composition. We consider that the attack is successful if the adversary is able to identify the circuit (see Threat model, 2nd bullet). Breaking the key is not necessary for this attack.

\begin{figure}[tb]
\centering 
\includegraphics[width=0.98\linewidth]{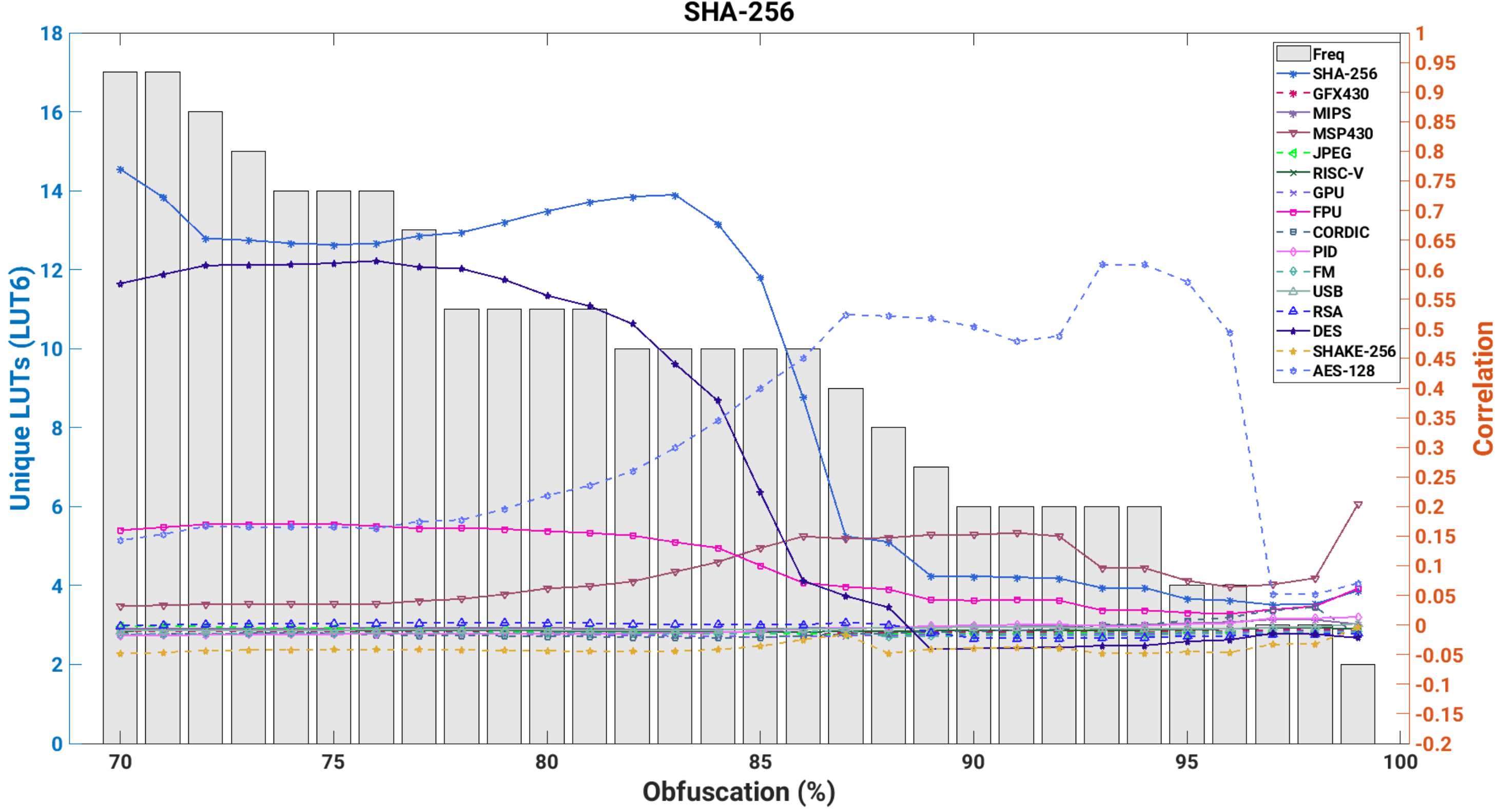}
\caption{The correlation of SHA-256 versus numerous other designs.}
\centering
\label{fig:corr_sha}
\end{figure}

In this experiment, we performed correlation analysis for the well-known SHA-256 crypto core as shown in Fig. \ref{fig:corr_sha}. The objective of this experiment is to analyze the leaked information from the static part of an obfuscated design against a database\footnote{We assume the adversary can obtain samples of open source cores from repositories and execute FPGA synthesis on them with his tool of choice.} of circuits that are known to the attacker. We have obfuscated SHA-256 in the 70-100\% range and then correlated the static portion of the design with the database of known circuits. In Fig. \ref{fig:corr_sha}, we show the results where the x-axis shows the obfuscation percentage and the y-axis shows correlation (right) and number of unique LUTs (left). For this circuit, three regions of interest can be defined: 97-100\% (no correlation), 86-96\% (strong correlation to another circuit), and 70-85\% (correlation to itself). 

This attack reveals that if the adversary goal is solely to identify the circuit's intent, eASIC can be as vulnerable as an ASIC design. To mitigate this undesirable effect, obfuscation levels should remain relatively high. Otherwise, if the obfuscation lies between 70 and 84\%, the search space would shift from L\textsubscript{3} to L\textsubscript{4} as shown in Fig.~\ref{fig:search_space}. This is a key finding of our manuscript and will guide our design choices when implementing the physical synthesis of eASIC designs in the section that follows.

\section{Physical Synthesis for eASIC} \label{sec:physical_implementation}

\begingroup
\setlength{\tabcolsep}{2.0pt} 
\renewcommand{\arraystretch}{1.1} 
\begin{table*} [t]
\footnotesize \centering
\caption{Results for the implementation of SHA-256 for different obfuscation levels}
\label{tab:sha_impl}
\begin{tabular}{|p{1cm}|p{1.5cm}|p{0.9cm}|p{1.5cm}|p{1cm}|p{1cm}|p{1.7cm}|p{1.6cm}|p{0.8cm}|p{1cm}|p{1cm}|p{0.7cm}|p{1.5cm}|}
\hline 
\textbf{CAD Flow} & \textbf{Obfuscation } & \textbf{Density} & \textbf{Area ($\mu m^2$)} & \textbf{Freq. (MHz)} & \textbf{Leakage ($mW$)} & \textbf{Dynamic Power ($mW$)} & \textbf{Total Power ($mW$)} & \textbf{\# LUT} & \textbf{\# Buffer}  & \textbf{\# Comb.} & \textbf{\# Inv.} & \textbf{\# Sequential} \\ \hline
FPGA & 100\% & -- & -- & 77 & 158 & 33 & 191 & 2238 & -- & -- & -- & 1830  \\ \hline
TOTe & 100\% & 46\% & 1412227.08 & 166.7 & 17.01 & 257.23 & 274.24 & 2238 & 6234 & 82756 & 4686 & 105128  \\ \hline
TOTe & 90\%  & 45\% & 1274690.16 & 178.6 & 15.34 & 246.87 & 262.21 & 2015 & 5411 & 83452 & 4188 & 94876 \\ \hline
TOTe & 85\%  & 46\% & 1215328.32 & 200   & 14.62 & 263.20 & 277.82 & 1904 & 5249 & 79626 & 3972 & 90420 \\ \hline
TOTe & 80\%  & 54\% & 1135752.20 & 200   & 13.93 & 248.84 & 262.77 & 1792 & 7699 & 74000 & 3755 & 83790 \\ \hline
ASIC & 0\%   & 71\% &   43097.40 & 200   & 0.525 & 6.405  & 6.93   &    0 &  119 &  3165 &  128 &  1806  \\ \hline
ASIC & 0\%   & 71\% &   60563.52 & 769   & 0.862 & 32.69  & 33.55   &    0 &  336 &  3165 &  128 &  1806  \\\hline
\end{tabular}
\end{table*}
\endgroup

This section contains the physical implementation results for an obfuscated SHA-256 core. We have selected SHA-256 as it is popular and widely used in cryptography. The variants of the design with different obfuscation levels are implemented with the aid of the LUTs defined in Section \ref{subsec:lut_implementation}. The results obtained after implementation are focused on performance vs. area trade-offs for the 80-100\% obfuscation range as determined by the security analysis of Fig. \ref{fig:corr_sha}. Initially, we synthesized and implemented the SHA-256 core on FPGA. This implementation achieves a frequency of only 77 MHz (for reference, the Kintex-7 family is produced on a 28nm CMOS technology). To start the analysis, we select 100\% obfuscation as a baseline design because it is fully reconfigurable and somewhat analogous to an FPGA design. 

The implementation results for 0\%, 80\%, 85\%, 90\%, and 100\% obfuscation are listed in Table \ref{tab:sha_impl}. These are obtained after physical synthesis and are for the worst process corner (SS) and a nominal temperature of $25^{\circ}$C. It is noteworthy that the performance of the design is increasing as we decrease the level of security. This behaviour is clearly depicted in the fourth column of Table \ref{tab:sha_impl} and matches the goal we set from the beginning: to trade performance for security. Here we also show that performance saturates rather quickly, as predicted by TOTe in Section \ref{subsec:design_space}. The area of the design is proportional to the obfuscation level which means that increasing the security of design will cause area overhead. As we only exploit LUT primitives for promoting obfuscation, the number of LUTs increases with the obfuscation level. In the same manner, leakage and dynamic power figures are proportional to security as reconfigurable logic is less efficient than static logic. The results obtained from the physical synthesis justify trade-offs and the last five columns of Table \ref{tab:sha_impl} show the resource requirements.

\begin{figure*}[t]
\centering \footnotesize
\includegraphics[width=0.95\linewidth]{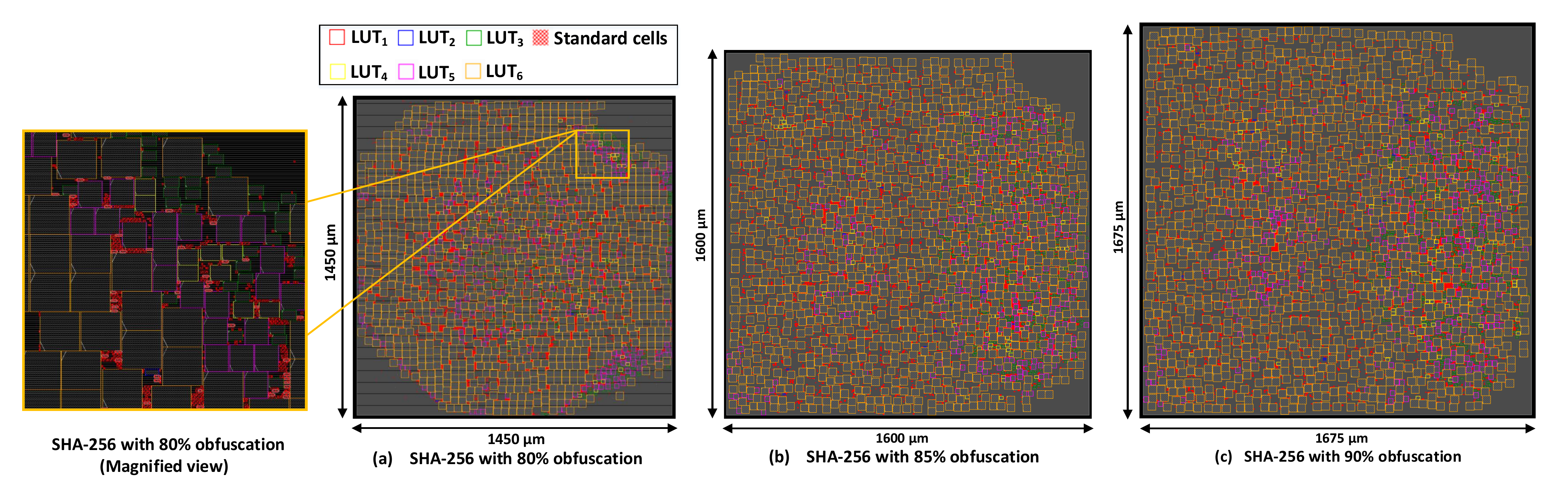}
\caption{Implementation results for SHA-256 with different obfuscation levels.}
\centering
\label{fig:obfuscation_impl}
\end{figure*}

 The first three panels (a, b and c) of Fig. \ref{fig:obfuscation_impl} illustrate the layouts for 80\%, 85\% and 90\% obfuscation levels. The dimensions of the layouts are indicated on the bottom and left sides of each panel. All the six variants of LUTs are highlighted with different colors and the static logic part of eASIC is highlighted in red -- notice that, as expected, the design remains primarily a sea of LUTs - the reconfigurable logic part. The majority of those LUTs are LUT\textsubscript{6}, thus the layouts appear to be dominated by yellow boxes. 
The placement of LUTs is done by the ASIC placement tool. For this, we modified the LUT macros to behave as regular standard cells. Then, the placer exploits its optimization strategies to place each LUT efficiently. From a magnified view, the mixed structure of LUT macros and standard cells clearly depicts the placement pattern, and that spaces between macros are usually filled with standard cells (static logic part). Notice how the LUT macros align with the standard cell rows, allowing for the entire design to have a uniform power rail and power stripe configuration. 

\section{Comparison and discussion} \label{sec:discussion}

From the many results, we conclude that obfuscation levels should be relatively high to achieve a considerable security, thus the majority of the eASIC logic should be reconfigurable logic (i.e., LUTs). Having a large portion of reconfigurable logic provides an opportunity to correct the issues/bugs that could be easily fixed during the reconfiguration phase. Naturally, there are limitations since a portion of the system consists of static logic and cannot be modified. This limitation could be eased if the eASIC layout were to include spare LUTs. These same spare LUTs could serve as decoy LUTs for preventing the composition attack described in Section \ref{subsec:composition_analysis}, but at a cost in area, power, and likely timing as well. To some degree, those spare LUTs could also be used to make side-channel attacks less successful. These possibilities are not studied in this manuscript but remain promising concepts for future work. 

 Our eASIC device presents a largely regular structure upon visual inspection. This effect can be modulated if it proves to be effective against a reverse engineering adversary. For instance, we could have mapped LUTs of all sizes to LUT\textsubscript{6}, which would increase the layout regularity (i.e., making it only red and yellow). Similarly, LUTs could have been laid out in a perfect grid fashion. These two design choices are relatively simple to implement in physical synthesis but carry overheads that we deemed not advantageous, even if they make perfect sense for an FPGA device.

A recent trend in obfuscation research is the use of embedded FPGA (eFPGA) \cite{e-FPGA1, e-FPGA2}. A very similar approach is also found in \cite{e-FPGA3}, where authors perform obfuscation with transistor-level granularity. While there are advantages to this practice, it has been used selectively to only protect key portions of a design and therefore keep the performance penalty as low as possible. The challenge is in determining which portions of the circuit merit protection and which ones do not. Our eASIC approach bypasses this question almost completely by only revealing (portions of) critical paths when they are selected to become static logic, which we consider an advantage if the ASIC-equivalent performance can be sacrificed. In \cite{re_2021}, the authors present a top-down methodology to implement ASICs with eFPGAs. Their designs share many of the advantages of our eASIC solution while presenting more regularity than our designs (they make use of logic tiles as in commercial FPGAs). Our tile-free design trades this regularity for performance as evidenced by the layout in Fig. \ref{fig:obfuscation_impl}.

\section{Conclusions} \label{sec:conclusion}
In this paper, we have developed a custom tool (TOTe) that obfuscates a design and transforms it into an eASIC device. Our eASIC solution contrasts with the current practice of eFPGA for obfuscation and this is not by coincidence: our experimental results show that obfuscation rates have to be high to protect not only the bitstream but also the design's intent. This is a key finding of our research which we hope can help to steer current obfuscation practices in the literature. Our findings are also validated in a commercial physical synthesis tool with industry-strength timing and power analysis, from which we confirm that TOTe's trade-off analysis is sufficiently accurate. Our future research will focus on the ideas put forward in our discussion, where we argue that eASIC has many benefits beyond obfuscation.

\bibliographystyle{IEEEtran}
\bibliography{obfuscation}


\end{document}